\documentclass[prb,preprint]{revtex4}
\pdfoutput=1
\begin{document}


\title{Comment on ``Analyzing collisions in classical mechanics using mass-momentum diagrams'' by A. Ogura, arXiv:1809.01768
\medskip }

\date{September 21, 2018} \bigskip

\author{Manfred Bucher \\}
\affiliation{\text{\textnormal{Physics Department, California State University,}} \textnormal{Fresno,}
\textnormal{Fresno, California 93740-8031} \\}

\begin{abstract}
A change of axes and notation would improve an otherwise excellent teaching diagram.
\end{abstract}

\maketitle

As the simplest process of particle interaction, a collision plays a basic role in physics. Although most collisions occur in two or three dimensions, a one-dimensional (``head-on'') collision serves as a preparatory step in physics education, illustrating conservation of momentum. It is frequently demonstrated in the classroom with gliders on an air track and likewise experimented with in the lab. Related problems occur in homework and exams. While necessary for calculations, the corresponding equations are too complicated to provide an immediate qualitative view of the situation before and after the collision. For this purpose ``situation diagrams'' of one-dimensional collisions have been proposed\cite{1,2} and for a practical application thereof.\cite{3}
\textbf{}

The diagram of Ref. 1 illustrates conservation of both momentum and (total) energy of the particles. However, the necessary diagram rules may be too complicated for easy recall. The new diagram by Ogura\cite{2} expresses conservation of momentum, assuming a given coefficient of restitution, \textit{e}. The diagram represents each colliding particle \textit{j}  by a state vector $\overrightarrow{\rm{\varepsilon}}_j = (m_j, p_j)$ in a rectangular \textit{m} vs. \textit{p} diagram. The diagram rules are simple, amounting essentially to vector addition in a plane. This provides an immediate overview of the situation. The purpose of this comment is to suggest improvements by (1) a change of axes to a Cartesian \textit{p} vs. \textit{m} diagram instead of the Ogura's \textit{m} vs. \textit{p} diagram, and (2) a change of notation for the state vectors to
$\overrightarrow{\rm{s_j}} = (m_j, p_j)$. There seems to be no stringent reason for the Ogura's choice of \textit{p} on the abscissa and \textit{m} on the ordinate or his use of the symbol $\overrightarrow{\rm{\varepsilon}}$ to prevent such changes.

The proposed modifications would improve the otherwise excellent diagram by three advantages: (i) The speed $v_j$ of particle \textit{j} is then shown by the slope of the corresponding vector in the Cartesian \textit{p} vs. \textit{m} diagram---in agreement with convention, by the tangent of the angle from the abscissa. Since speed is a directly observable quantity, in contrast to the composed quantity of momentum, $p_j = m_j v_j$, this feature of the diagram provides additional insight. (ii) The proposed \textit{p} vs. \textit{m} diagram would be consistent with the Ogura's notation of state vectors $(m_j, p_j)$ for particles \textit{j}, comparable to a vector $(x,y)$ in the common Cartesian \textit{y} vs. \textit{x} diagram. (iii) The letter `\textit{s}' of the vector $\overrightarrow{\rm{s}}$ would mnemonically indicate the ``state'' of the collision better than the letter $\varepsilon$ of the vector $\overrightarrow{\rm{\varepsilon}}$.

\end{document}